\documentclass[]{article}

\usepackage{fullpage}
\usepackage{authblk}
\usepackage{amsfonts}
\usepackage{multirow}
\usepackage{mathrsfs}
\usepackage{graphicx}
\usepackage{amsmath}
\usepackage{amssymb}
\usepackage{bm}
\usepackage{bbm}
\usepackage{color}
\usepackage{slashed}
\usepackage{diagbox}
\usepackage{ulem}
\usepackage{wrapfig}
\usepackage[hidelinks]{hyperref}
\usepackage{cleveref}
\usepackage{fancyhdr}
\usepackage{verbatim}
\numberwithin{equation}{section}

\begin{document}
%%%%%%%%%%%%%%%%%%%%%%%%%%%%%%%%%%%%%%%%%%%%%%%%%%%%%%%%%%%%%%%%%%%%%%%%%%%%%%

\fancypagestyle{plain}{
    \fancyhead[R]{Alberta Thy 16-19}
    \renewcommand{\headrulewidth}{0pt}
}

\title{\bf{\textsf{Reduction of Feynman Integrals in the Parametric Representation II: Reduction of Tensor Integrals}}}

\author[a]{\bf{\textsf{Wen Chen}}\footnote{wchen1@ualberta.ca}}
\affil[a]{\small{\it{Department of Physics, University of Alberta, Edmonton, AB, T6G 2E1, Canada}}}

\date{\small{\today}}

\maketitle

%%%%%%%%%%%%%%%%%%%%%%%%%%%%%%%%%%%%%%%%%%%%%%%%%%%%%%%%%%%%%%%%%%%%%%%%%%%%%%
\begin{abstract}
In a recent paper by the author (Chen in JHEP 02:115, 2020), the reduction of Feynman integrals in the parametric representation was considered. Tensor integrals were directly parametrized by using a generator method. The resulting parametric integrals were reduced by constructing and solving parametric integration-by-parts (IBP) identities. In this paper, we furthermore show that polynomial equations for the operators that generate tensor integrals can be derived. Based on these equations, two methods to reduce tensor integrals are developed. In the first method, by introducing some auxiliary parameters, tensor integrals are parametrized without shifting the spacetime dimension. The resulting parametric integrals can be reduced by using the standard IBP method. In the second method, tensor integrals are (partially) reduced by using the technique of Gr\"obner basis combined with the application of symbolic rules. The unreduced integrals can further be reduced by solving parametric IBP identities.

\end{abstract}
%%%%%%%%%%%%%%%%%%%%%%%%%%%%%%%%%%%%%%%%%%%%%%%%%%%%%%%%%%%%%%%%%%%%%%%%%%%%%%
%%\pacs{\it}

\tableofcontents

\section{Introduction}

One of the most important issues in today's high energy physics is the calculation of Feynman amplitudes. Generally, Feynman amplitudes are expressed in terms of tensor integrals, that is, Feynman integrals with Lorentz indices. A widely used strategy to calculate tensor integrals is to first do a tensor reduction to write tensor integrals as linear combinations of scalar integrals, and then reduce the scalar integrals by using the integration-by-parts (IBP) method \cite{Tkachov:1981wb,Chetyrkin:1981qh,Laporta:2001dd}.

For one-loop integrals, the tensor reduction can be implemented by the well-known Passarino-Veltman reduction \cite{Passarino:1978jh}. Multiloop tensor integrals can be reduced to scalar integrals by using the projector technique (see e.g. ref. \cite{Garland:2002ak}). Specifically, an amplitude is written as a linear combination of some tensor structures. The coefficient of a tensor structure is extracted by applying a projector to the amplitude. In practice, the tensor structures and the corresponding projectors are process-dependent. A general algorithm can be developed in principle. However, the calculation of the projectors for high-rank tensor integrals, which involves the inversion of a large matrix, is quite cumbersome. Recently, in refs. \cite{Chen:2019wyb,Peraro:2019cjj} it was shown that due to the four-dimensional nature of the external states the number of projectors could be much smaller by considering helicity amplitudes. These methods explicitly make use of the structures of the amplitudes for the process under consideration. Hence they break down for a general tensor integral.

An alternative to tensor reduction is to directly parametrize tensor integrals, as was suggested in ref. \cite{Tarasov:1996br}. It was suggested in ref. \cite{Lee:2014tja} that it was possible to derive IBP relations directly in the Lee-Pomeransky representation \cite{Lee:2013hzt}. Parametric integrals could be reduced to master integrals by solving these linear relations. In this paper, we follow a similar approach developed in ref. \cite{Chen:2019mqc} (referred to as paper I hereafter). A tensor integral is parametrized by applying a sum of chains of index-shifting operators to a scalar integral. The advantage of this method is that the parametrization can be applied to arbitrary tensor integrals, and can easily be implemented in automatic calculations. The resulting parametric integrals can directly be reduced by solving parametric IBP identities. As was mentioned in paper I,  the reduction in the parametric representation was advantageous over the traditional IBP method in several aspects. Comparing with the momentum-space IBP method, fewer IBP identities are generated for a reduction. Symbolic rules can easily be generated in the parametric representation. \footnote{Symbolic rules can also be generated in the regular momentum-space IBP method \cite{Lee:2012cn}. However, this has to be done sector by sector, which is much more complicated.} Furthermore, symmetries of integrals (under permutations of indices) are more transparent in the parametric representation.

The drawback of this approach is that the highest degree (to be specified in \cref{Sec:LinRelParInt}) of the parametric integrals increases fast with the tensor ranks. Consequently, the number of linear relations to be solved increases rapidly. In this paper, we will show that this problem can be solved by directly constructing and solving polynomial equations for operators that generate tensor integrals. Based on these equations, two approaches to reduce tensor integrals are provided.

This paper is organized as follows. In \cref{Sec:LinRelParInt}, we give a brief review of the method developed in paper I. In \cref{Sec:RedTensInt}, we described the two methods to reduce tensor integrals. Some examples are provided in \cref{Sec:Exampl}.

\section{Linear relations between parametric integrals}\label{Sec:LinRelParInt}

We consider a $L$-loop integral

\begin{equation}\label{Eq:MomSpacInt}
M=\pi^{-\frac{1}{2}Ld}\int d^dl_1d^dl_2\cdots d^dl_L\frac{1}{D_1^{\lambda_1+1}D_2^{\lambda_2+1}\cdots D_n^{\lambda_n+1}},
\end{equation}

\noindent where $d$ is the dimensionally regularized spacetime dimension, and $D_i$ are inversed propagators. We define $\sum_{i=1}^nx_iD_i\equiv\sum_{i,j=1}^LA_{ij}l_i\cdot l_j+2\sum_{i=1}^LB_i\cdot l_i+C$. Following the convention used in paper I, this integral is parametrized by the integral

\begin{equation}
\begin{split}
I(\lambda_0,\lambda_1,\dots,\lambda_n)=&s_g^{L/2}e^{-i\pi\lambda_f}M\\
=&\frac{\Gamma(-\lambda_0)}{\prod_{i=1}^{n+1}\Gamma(\lambda_i+1)}\int d\Pi^{(n+1)}\mathcal{F}^{\lambda_0}\prod_{i=1}^{n+1}x_i^{\lambda_i}\equiv\int d\Pi^{(n+1)}\mathcal{I}^{(-n-1)},\quad \lambda\notin\mathbb{Z}^-.
\end{split}
\end{equation}

\noindent Here $s_g$ is the determinant of the dimensionally regularized spacetime metric and $\lambda_f\equiv\frac{1}{2}dL-n-\sum_{i=1}^n\lambda_i$. $\lambda_0$ is related to the spacetime dimension through $\lambda_0=-\frac{d}{2}$. The measure $d\Pi^{(n)}\equiv \prod_{i=1}^{n+1}dx_i\delta(1-\sum_j|x_j|)$, where the sum in the delta function runs over any nontrivial subset of $\{x_1,x_2,\ldots,x_{n+1}\}$. $\mathcal{F}$ is a homogeneous polynomial of $x_i$ of degree $L+1$, defined by $\mathcal{F}(x)\equiv F(x)+U(x)x_{n+1}$. $U$ and $F$ are Symanzik polynomials, defined by $U(x)\equiv\det{A}$, and $F(x)\equiv U(x)\left(\sum_{i,j=1}^L(A^{-1})_{ij}B_i\cdot B_j-C\right)$. In this paper, we define the degree of a parametric integral by $\Lambda\equiv\sum_{i=1}^{n}\lambda_i$.

As is proven in paper I, the parametric integral satisfies the following identities.

\begin{equation}\label{Eq:IBP1}
0=\int d\Pi^{(n+1)}\frac{\partial}{\partial x_i}\mathcal{I}^{(-n)}+\delta_{\lambda_i0}\int d\Pi^{(n)}\left.\mathcal{I}^{(-n)}\right|_{x_i=0},\qquad i=1, 2,\ldots, n+1,~\lambda\notin\mathbb{Z}^-,
\end{equation}

\noindent where $\delta_{\lambda_i0}$ is the Kronecker delta. We define the index-shifting operators $R_i$, $D_i$, and $A_i$, with $i=0,1,\dots,n$, such that

\begin{subequations}
\begin{align}
R_iI(\lambda_0,\dots,\lambda_i,\dots,\lambda_n)=&(\lambda_i+1)I(\lambda_0,\dots,\lambda_i+1,\dots,\lambda_n),\\
D_iI(\lambda_0,\dots,\lambda_i,\dots,\lambda_n)=&I(\lambda_0,\dots,\lambda_i-1,\dots,\lambda_n),\\
A_iI(\lambda_0,\dots,\lambda_i,\dots,\lambda_n)=&\lambda_iI(\lambda_0,\dots,\lambda_i,\dots,\lambda_n),\quad\lambda\notin\mathbb{Z}^-.
\end{align}
\end{subequations}

\noindent It is understood that

\begin{subequations}
\begin{align}
&I(\lambda_0,\dots,\lambda_{i-1},-1,\dots,\lambda_n)\equiv\int d\Pi^{(n)}\left.\mathcal{I}^{(-n)}\right|_{x_i=0},\quad\text{and}\\
&R_iI(\lambda_0,\dots,\lambda_{i-1},-1,\dots,\lambda_n)\equiv0.
\end{align}
\end{subequations}

\noindent The product of two operators are defined by successive action. That is, $(XY)I(\lambda)\equiv X(YI(\lambda))$. It is easy to get the following commutation relations:

\begin{align}
D_iR_j-R_jD_i=&\delta_{ij},\\
D_iA_j-A_jD_i=&\delta_{ij}D_i,\\
R_iA_j-A_jR_i=&-\delta_{ij}R_i.
\end{align}

\noindent We formally define operators $D_{n+1}$ and $R_{n+1}$, such that $D_{n+1}I=I$, and $R_{n+1}^iI=(A_{n+1}+1)(A_{n+1}+2)\cdots(A_{n+1}+i)I$, with $A_{n+1}\equiv-(L+1)A_0-\sum_{i=1}^n(A_i+1)$. Notice that $R_{n+1}$, $D_{n+1}$, and $A_{n+1}$ do not obey the commutation relations listed above. It is easy to rewrite \cref{Eq:IBP1} in the following form:

\begin{equation}\label{Eq:IBP2}
D_0\frac{\partial\mathcal{F}(R)}{\partial R_i}-D_i\approx 0,\quad i=1,~2,\dots,n+1.
\end{equation}

\noindent Here we use ``$\approx$'' instead of ``$=$'' to indicate that these identities are valid only when they are applied to nontrivial parametric integrals $I(\lambda)$. It should be noted that $X_i0$ is ill-defined for $X_i=R_i,D_i,\text{ or }A_i$. Thus $R_i$, $D_i$, and $A_i$ are not linear operators on the linear space of parametric Feynman integrals. Notice that $R_{n+1}$ does not commute with $R_i$ for $i=0,1,\dots,n$. We assume that $R_{n+1}$ in $\mathcal{F}(R)$ is to the right of $U(R)$. Explicitly, we have

\begin{subequations}\label{Eq:IBP3}
\begin{align}
0&\approx D_0U(R)-1,\label{Eq:IBP3:a}\\
0&\approx D_0\left(\frac{\partial F(R)}{\partial R_i}+\frac{\partial U(R)}{\partial R_i}R_{n+1}\right)-D_i.\label{Eq:IBP3:b}
\end{align}
\end{subequations}

\noindent Obviously, for two functions $f$ and $g$, $f\approx0$ implies $fg\approx0$. By using the homogeneity of $U$ and $F$, it can be proven that

\begin{equation}\label{Eq:IBP4}
D_0\mathcal{F}+A_0-1\approx0.
\end{equation}

Due to the $D_0$-dependence, \cref{Eq:IBP2} may shift the spacetime dimension \cite{Tarasov:1996br,Laporta:2001dd}. Identities free of dimensional shift can be obtained by using a method similar to the parametric-annihilator method \cite{Baikov:1996iu,Lee:2013hzt,Lee:2014tja,Bitoun:2017nre} or the syzygy-equation method \cite{Larsen:2015ped}. Specifically, let $f_i$ be a list of polynomials in the $R$ of degree $\gamma$, satisfying the following equations:

\begin{subequations}
\begin{align}
\sum_{i=1}^nf_i\frac{\partial U}{\partial R_i}=&a_1U+b_1F,\\
\sum_{i=1}^nf_i\frac{\partial F}{\partial R_i}=&a_2U+b_2F,
\end{align}
\end{subequations}

\noindent where $a$ and $b$ are also polynomials in the $R$. By virtue of \cref{Eq:IBP3,Eq:IBP4}, we have

\begin{equation}\label{Eq:DimShiftFreeIBP}
\begin{split}
0\approx&\sum_{i=1}^n\left\{D_0\left[\frac{\partial F(R)}{\partial R_i}+\frac{\partial U(R)}{\partial R_i}(A_{n+1}+1)\right]-D_i\right\}f_i\\
=&D_0\sum_{i=1}^n\frac{\partial F(R)}{\partial R_i}f_i+D_0\sum_{i=1}^n\frac{\partial U(R)}{\partial R_i}f_i(A_{n+1}-\gamma+1)-\sum_{i=1}^nD_if_i\\
=&D_0\mathcal{F}\left[b_2+(A_{n+1}-1)b_1\right]+D_0U\left[a_2-(A_{n+1}+1)b_2+A_{n+1}a_1-(A_{n+1}^2-1)b_1\right]-\sum_{i=1}^nD_if_i\\
\approx&-A_{n+1}^2b_1-A_0A_{n+1}b_1+A_{n+1}(a_1+b_1-b_2)+A_0(b_1-b_2)+a_2-\sum_{i=1}^nD_if_i.
\end{split}
\end{equation}

\noindent Obviously the last line in the above equation is free of dimensional shift. In practical calculations, symbolic rules can be derived by solving these identities. We prescribe that $R_i$ is of degree $1$, $D_i$ is of degree $-1$, and $A_i$ is of degree $0$. Then in the case of the absence of negative indices (as is the case for Method II to be described in the next section), the ordering of monomials in \cref{Eq:DimShiftFreeIBP} is consistent with the ordering of the corresponding parametric integrals. Symbolic rules can easily be generated out of \cref{Eq:DimShiftFreeIBP} in this case. For example, for the tadpole integral with a mass $m$, we have a symbolic IBP identity $m^2R_1-(A_0+A_1+1)\approx0$. According to the ordering, $R_1$ is superior to $A_0$ and $A_1$, so this equation is solved by $R_1\approx\frac{1}{m^2}(A_0+A_1+1)$. Correspondingly we have the symbolic rule $I(-\frac{d}{2},i)=\frac{2i-d}{2m^2i}I(-\frac{d}{2},i-1)$.

\section{Reduction of tensor integrals}\label{Sec:RedTensInt}

According to the derivation in paper I, a tensor integral can be parametrized by\footnote{This expression is consistent with the parametrization used in ref. \cite{Heinrich:2008si}.}

\begin{equation}\label{Eq:ParTensInt}
\begin{split}
M_{i_1i_2\cdots i_r}^{\mu_1\mu_2\cdots\mu_r}\equiv&\pi^{-\frac{1}{2}Ld}\int d^dl_1d^dl_2\cdots d^dl_L\frac{l_{i_1}^{\mu_1}l_{i_2}^{\mu_2}\cdots l_{i_r}^{\mu_r}}{D_1^{\lambda_1+1}D_2^{\lambda^2+1}\cdots D_n^{\lambda_n+1}}\\
=&s_g^{-L/2}e^{i\pi\lambda_f}\left[P_{i_1}^{\mu_1}P_{i_2}^{\mu_2}\cdots P_{i_r}^{\mu_r}I(-\frac{d}{2},\lambda_1,\lambda_2,\ldots,\lambda_n)\right]_{p^\mu=0},
\end{split}
\end{equation}

\noindent where the operator

\begin{equation}\label{Eq:TensGen}
P_i^\mu(p)\equiv-\frac{\partial}{\partial p_{i,\mu}}-\widetilde{B}_i^\mu+\frac{1}{2}\sum_{j=1}^L\widetilde{A}_{ij}p_j^\mu,
\end{equation}

\noindent where $\widetilde{A}_{ij}\equiv D_0U(A^{-1})_{ij}$ and $\widetilde{B}_i^\mu\equiv\sum_{j=1}^L\widetilde{A}_{ij}B_j^\mu$. Thus a tensor integral is parametrized by a linear combination of parametric integrals of the form

\begin{equation}\label{Eq:TensParInt}
M_i=f_i(\widetilde{A},\widetilde{B})I(\lambda_0,\lambda_1,\dots,\lambda_n).
\end{equation}

Obviously $\widetilde{A}(x)$ is of degree $L-1$ in $x$, and $\widetilde{B}(x)$ is of degree $L$ in $x$. Due to the $\widetilde{B}$ term in \cref{Eq:TensGen}, the highest degree of parametric integrals for a $L$-loop rank-$r$ tensor integral is $Lr$, which increases rapidly with $r$ for multiloop integrals. Consequently, the linear system to be solved is very large for a high-rank tensor integral. A solution to this problem is to directly reduce $f_i$ in \cref{Eq:TensParInt} without substituting the explicit forms of the $\widetilde{B}$.\footnote{The $\widetilde{A}$ term is less problematic, so we do not need to pay much attention to it.} As is derived in \cref{App:DerEqIBPB}, the $\widetilde{B}$ satisfy the following equations:

\begin{equation}\label{Eq:IBPB}
\sum_{j,k=1}^L\frac{\partial A_{jk}}{\partial R_i}\widetilde{B}_j\cdot\widetilde{B}_k-2\sum_{j=1}^L\frac{\partial B_j}{\partial R_i}\cdot\widetilde{B}_j+D_0A_0\frac{\partial U}{\partial R_i}+\frac{\partial C}{\partial R_i}+D_i\approx0.
\end{equation}

\noindent Based on these equations, we trade the reduction of the $f_i$ in \cref{Eq:TensParInt} to a problem of polynomial reduction. The difficulty is that in \cref{Eq:IBPB} $D$'s don't commute with the $\widetilde{B}$. We provide two methods to solve this problem, as will be described in the next two subsections.

\subsection{Method I}\label{Sec:MethI}

In this subsection, by introducing some auxiliary parameters, we will show that the $\widetilde{B}$ can be expressed as linear combinations of the $D$, which are of degree $-1$ (according to the prescription at the end of \cref{Sec:LinRelParInt}). Consequently, the corresponding parametric integrals are of lower degrees. First, we need to extend the definition of the parametric integrals to include integrals with negative indices. We define \cite{Lee:2014tja}

\begin{equation}
\begin{split}
&I(\lambda_0,\lambda_1,\ldots,\lambda_{i-1},-m,\lambda_{i+1},\ldots,\lambda_n)\\
\equiv&\lim_{\lambda_i\to-m}I(\lambda_0,\lambda_1,\ldots,\lambda_{i-1},\lambda_i,\lambda_{i+1},\ldots,\lambda_n)\\
=&(-1)^{m-1}\frac{\Gamma(-\lambda_0)}{\prod_{j\neq i}^{n+1}\Gamma(\lambda_j+1)}\int d\Pi^{(n)}\left[\frac{\partial^{m-1}\mathcal{F}^{\lambda_0}}{\partial x_i^{m-1}}\right]_{x_i=0}\prod_{j\neq i}^{n+1}x_j^{\lambda_j},\quad m\in N.
\end{split}
\end{equation}

\noindent It is easy to see that this definition is consistent with \cref{Eq:IBP2}. The definition of the degree of an integral should also be modified. We define the degree of a nonnegative index $\lambda_i$ by $d_i\equiv\lambda_i$, and the degree of a negative index $\lambda_i$ by $d_i\equiv-1-\lambda_i$. The degree of an integral is defined by $\Lambda\equiv\sum_{i=1}^n d_i$.

Let $a_{ij}$ and $b_{ij}$ be the solutions of

\begin{subequations}\label{Eq:abDef}
\begin{align}
\sum_jb_{ij}\frac{\partial A(R)}{\partial R_j}=&0,\\
\sum_ja_{ij}\frac{\partial B(R)}{\partial R_j}=&0,
\end{align}
\end{subequations}

\noindent where $A$ and $B$ are defined in \cref{Sec:LinRelParInt} (below \cref{Eq:MomSpacInt}). Solutions of these two equations may be linearly dependent. We denote those linearly dependent solutions by $c_{ij}$. By default, we assume that these solutions are excluded from $a_{ij}$ and $b_{ij}$. For brevity, we denote

\begin{subequations}
\begin{align}
\frac{\partial}{\partial a_i}\equiv\sum_ja_{ij}\frac{\partial}{\partial R_j},\qquad&D_{a_i}\equiv\sum_ja_{ij}D_j,\\
\frac{\partial}{\partial b_i}\equiv\sum_jb_{ij}\frac{\partial}{\partial R_j},\qquad&D_{b_i}\equiv\sum_jb_{ij}D_j,\\
\frac{\partial}{\partial c_i}\equiv\sum_jc_{ij}\frac{\partial}{\partial R_j},\qquad&D_{c_i}\equiv\sum_jc_{ij}D_j.
\end{align}
\end{subequations}

\noindent $B_i^\mu$ is of the form $\sum_uB_{iu}Q_u^\mu$, where $Q_u^\mu$ are the (linearly independent) external momenta. Similarly we have $\widetilde{B}_i^\mu=\sum_u\widetilde{B}_{iu}Q_u^\mu$. By virtue of \cref{Eq:IBPB}, we have

\begin{subequations}
\begin{align}
-2\sum_{u,v,j}Q_u\cdot Q_v\frac{\partial B_{ju}}{\partial b_i}\widetilde{B}_{jv}+\frac{\partial C}{\partial b_i}+D_{b_i}\approx&0,\label{Eq:IBPBb}\\
\sum_{j,k}\frac{\partial A_{jk}}{\partial a_i}\widetilde{B}_{j}\cdot\widetilde{B}_{k}+D_0A_0\frac{\partial U}{\partial a_i}+\frac{\partial C}{\partial a_i}+D_{a_i}\approx&0,\label{Eq:IBPBa}\\
\frac{\partial C}{\partial c_i}+D_{c_i}\approx&0.\label{Eq:IBPBc}
\end{align}
\end{subequations}

Generally speaking, the matrix $\frac{\partial B_{ju}}{\partial b_i}$ is not invertible. However, we can always make it invertible by introducing some auxiliary parameters $x_{n+1},~x_{n+2},\dots$ through the transformation $B_{iu}\to B_{iu}+\sum_k c_kx_k$, where $c_k$ are some constants. This is equivalent to adding some auxiliary propagators  of the form $\sum c_{ij}l_i\cdot Q_j$. We denote the inverse of $\frac{\partial B_{ju}}{\partial b_i}$ by $\beta_{ju,i}$. That is,

\begin{equation}
\sum_k\beta_{iu,k}\frac{\partial B_{jv}}{\partial b_k}=\delta^{ij}\delta^{uv}.
\end{equation}

\noindent Similarly, we can always make the Gram matrix $Q_u\cdot Q_v$ invertible by introducing some auxiliary external momenta. (Notice that we assume that the $Q$ are linearly independent.) Then \cref{Eq:IBPBb} is solved by

\begin{equation}\label{Eq:BtSol}
\widetilde{B}_{iu}\approx\frac{1}{2}\sum_{j,v}g_{uv}\beta_{iv,j}\left(\frac{\partial C}{\partial b_j}+D_{b_j}\right)\equiv\bar{B}_{iu},
\end{equation}

\noindent where $g_{uv}$ is the inverse of $Q_u\cdot Q_v$.

Since the $\widetilde{B}$ are of degree $L$, while the $D$ (hence the $\bar{B}$) are of degree $-1$, by expressing $\widetilde{B}_{iu}$ in terms of $\bar{B}_{iu}$, the corresponding parametric integrals are of much lower degrees. However, we cannot directly apply \cref{Eq:BtSol} to a chain of the $\widetilde{B}$, because the $\bar{B}$ do not commute with the $\widetilde{B}$. It is easy to get the commutation relations

\begin{subequations}
\begin{align}
\left[\bar{B}_{iu},\widetilde{B}_{jv}\right]=&\frac{1}{2}g_{uv}\widetilde{A}_{ij},\label{Eq:BtBbComm1}\\
\left[\bar{B}_i^\mu,\widetilde{B}_j^\nu\right]=&\frac{1}{2}\sum_{u,v}g_{uv}Q_u^\mu Q_v^\nu\widetilde{A}_{ij}\equiv\frac{1}{2}\eta^{\prime \mu\nu}\widetilde{A}_{ij}.\label{Eq:BtBbComm2}
\end{align}
\end{subequations}

\noindent By virtue of these commutation relations, together with \cref{Eq:BtSol}, it can be shown that the operator $P_i^\mu$ defined in \cref{Eq:TensGen} can be replaced by (for the proof, see \cref{App:ProofEqTensGen2})

\begin{equation}\label{Eq:TensGen2}
P_i^\mu\approx-\frac{\partial}{\partial\bar{p}_{i,\mu}}-\bar{B}_i^\mu+\frac{1}{2}\sum_j\widetilde{A}_{ij}\bar{p}_j^\mu,
\end{equation}

\noindent where $\bar{p}_i$ are vectors such that $\bar{p}_i\cdot Q_j=0$. The $\bar{B}$ are free of $D_0$, and thus will not shift the spacetime dimension. Due to the definition of $\beta_{ij,k}$, the $\bar{B}$ commute with $\widetilde{A}$'s. Thus by applying operators $P_i^\mu$, tensor integrals are parametrized by integrals of the form $f(\widetilde{A})I(-\frac{d}{2},\ldots)$. It remains to reduce chains of $\widetilde{A}$.

Similar to $\frac{\partial B_{j}}{\partial b_i}$, the matrix $\frac{\partial A_{jk}}{\partial a_i}$ can be made invertible by introducing some auxiliary parameters through the transformation $A_{ij}\to A_{ij}+\sum_kc_kx_k$. This is equivalent to adding some auxiliary propagators of the form $\sum_{ij}c_{ij}l_i\cdot l_j$. Let $\alpha_{ij,k}$ be the matrix such that

\begin{equation}
\sum_k\alpha_{ij,k}\frac{\partial A_{mn}}{\partial a_k}=\frac{1}{2}\left(\delta^{im}\delta^{jn}+\delta^{in}\delta^{jm}\right).
\end{equation}

\noindent As is derived in \cref{App:DerEqAtSol}, \cref{Eq:IBPBa} is solved by

\begin{equation}\label{Eq:AtSol}
\widetilde{A}_{ij}\left(A_0+\frac{E}{2}\right)\approx-\bar{B}_i\cdot\bar{B}_j-\sum_k\alpha_{ij,k}\left(D_{a_k}+\frac{\partial C}{\partial a_k}\right)\equiv\bar{A}_{ij},
\end{equation}

\noindent where $\bar{B}_i^\mu\equiv\sum_u\bar{B}_{iu}Q_u^\mu$, and $E$ is the number of the external momenta. We have the following commutation relation:

\begin{equation}\label{Eq:AbAtComm}
\left[\bar{A}_{ij},\widetilde{A}_{kl}\right]\approx\frac{1}{2}\left(\widetilde{A}_{ik}\widetilde{A}_{jl}+\widetilde{A}_{il}\widetilde{A}_{jk}\right)-\widetilde{A}_{ij}\widetilde{A}_{kl}.
\end{equation}

\noindent By virtue of \cref{Eq:AtSol,Eq:AbAtComm}, we get

\begin{equation}\label{Eq:AtChainSol}
\begin{split}
\widetilde{A}_{i_2j_2}\widetilde{A}_{i_3j_3}\cdots\widetilde{A}_{i_nj_n}\bar{A}_{i_1j_1}\approx&\widetilde{A}_{i_1j_1}\widetilde{A}_{i_2j_2}\cdots\widetilde{A}_{i_nj_n}(A_0+\frac{E}{2})-\frac{1}{2}(\widetilde{A}_{i_1i_2}\widetilde{A}_{j_1j_2}+\widetilde{A}_{i_1j_2}\widetilde{A}_{i_2j_1})\widetilde{A}_{i_3j_3}\cdots\widetilde{A}_{i_nj_n}\\
&-\frac{1}{2}(\widetilde{A}_{i_1i_3}\widetilde{A}_{j_1j_3}+\widetilde{A}_{i_1j_3}\widetilde{A}_{i_3j_1})\widetilde{A}_{i_2j_2}\widetilde{A}_{i_4j_4}\cdots\widetilde{A}_{i_nj_n}\\
&-\cdots\\
&-\frac{1}{2}(\widetilde{A}_{i_1i_n}\widetilde{A}_{j_1j_n}+\widetilde{A}_{i_1j_n}\widetilde{A}_{i_nj_1})\widetilde{A}_{i_2j_2}\widetilde{A}_{i_3j_3}\cdots\widetilde{A}_{i_{n-1}j_{n-1}}\\
\equiv&\Delta_{j_1j_2\cdots j_n}^{i_1i_2\cdots i_n}.
\end{split}
\end{equation}

\noindent The proof of this equation can be found in \cref{{App:DerEqAtChainSol}}. A chain of $\widetilde{A}$ can further be reduced to a sum of chains of $\bar{A}$ by solving this equation. Currently we have not worked out the general solution to this equation yet. However, this is not a problem in practice, because this equation system is small in size and can easily be solved by brute force.

After applying \cref{Eq:TensGen2} and solutions of \cref{Eq:AtChainSol}, tensor integrals are parametrized by scalar integrals defined in dimension $d$. The resulting scalar integrals can be reduced by solving IBP identities free of dimensional shift, which can be obtained by multiplying both sides of \cref{Eq:BtSol,Eq:AtSol} with $A_{ij}$. Together with \cref{Eq:IBPBc}, we have

\begin{subequations}\label{Eq:DimShiftFreeIBPII}
\begin{align}
\frac{\partial C}{\partial c_i}+D_{c_i}\approx&0,\\
\sum_j\bar{B}_{ju}A_{ij}\approx&B_{iu},\\
\sum_k\bar{A}_{ik}A_{kj}\approx&(A_0+\frac{E}{2})\delta_{ij}.
\end{align}
\end{subequations}

\noindent These identities are nothing but the correspondences of the momentum-space IBP identities \cite{Baikov:1996iu} in the parametric representation \cite{Bitoun:2017nre}.

As a byproduct, \cref{Eq:BtSol} provides a method to construct differential equations in the parametric representation without shifting the spacetime dimension. Let $s$ be a kinematic variable, and assume that $A_{ij}$ is free of $s$. Then we have

\begin{equation}
\begin{split}
\frac{\partial}{\partial s}=&-D_0\frac{\partial\mathcal{F}}{\partial s}\\
=&-D_0\frac{\partial F}{\partial s}\\
=&-D_0U\left(\sum_{i,j,u,v}A^{-1}_{ij}B_{iu}B_{jv}\frac{\partial Q_u\cdot Q_v}{\partial s}+2\sum_{i,j,u,v}Q_u\cdot Q_vA^{-1}_{ij}B_{iu}\frac{\partial B_{jv}}{\partial s}-\frac{\partial C}{\partial s}\right)\\
\approx&-\sum_{i,u,v}\bar{B}_{iu}B_{iv}\frac{\partial Q_u\cdot Q_v}{\partial s}-2\sum_{i,u,v}Q_u\cdot Q_v\bar{B}_{iu}\frac{\partial B_{iv}}{\partial s}+\frac{\partial C}{\partial s}
\end{split}
\end{equation}

\subsection{Method II}\label{Sec:MethII}

In this subsection, we will show how to reduce tensor integrals without introducing auxiliary parameters by using the technique of Gr\"obner basis. In principle, we can generate a Gr\"obner basis \cite{Buchberger:1998aa} out of \cref{Eq:IBP3:a,Eq:IBPB}\footnote{Here \cref{Eq:IBP3:a} is considered because $D_0U$ contributes to $\widetilde{B}_i^\mu$. We notice that the $\widetilde{B}$ are not uniquely determined. It is easy to see that for a shift $\delta l_i^\mu$ of the loop momenta the corresponding shift of $\widetilde{B}_i^\mu$ is $\delta\widetilde{B}_i^\mu=D_0U\delta l_i^\mu$.}(For a brief introduction to Gr\"obner bases and relevant topics, see e.g. ref. \cite{Zhang:2016kfo}). Then the reduction of Feynman integrals is just a matter of polynomial reduction. The idea to use Gr\"obner bases to reduce Feynman integrals was first suggested by ref. \cite{Tarasov:1998nx}. The method of the s-basis \cite{Smirnov:2005ky,Smirnov:2006tz}, a variant of the Gr\"obner basis, was implemented in the early version of FIRE \cite{Smirnov:2008iw}. However, experience shows that generating a Gr\"obner basis for a noncommutative algebra is extremely time-consuming, which makes it less efficient for the reduction of Feynman integrals in practice. In this subsection, we try to solve this problem by converting the problem of a noncommutative algebra to the one of a commutative algebra. Though the Gr\"obner basis for the commutative algebra is not the full Gr\"obner basis for the corresponding noncommutative algebra, it can be used to greatly simply integrals to be reduced in practice.

Since $\frac{\partial A}{\partial R_i}$, $\frac{\partial B}{\partial R_i}$, and $\frac{\partial C}{\partial R_i}$ are constants, the l.h.s. of \cref{Eq:IBPB} is a polynomial of $\widetilde{B}_i^\mu$ and $D_i$, except for the $\frac{\partial U}{\partial R_i}$ term. By using \cref{Eq:IBP3} and the identity $D_if(R)-f(R)D_i=\frac{\partial f(R)}{\partial R_i}$, we can write \cref{Eq:IBPB} in the following form:

\begin{equation}\label{Eq:IBPB2}
\sum_{j,k=1}^L\frac{\partial A_{jk}}{\partial R_i}\widetilde{B}_j\cdot\widetilde{B}_k-2\sum_{j=1}^L\frac{\partial B_j}{\partial R_i}\cdot\widetilde{B}_j+\frac{\partial C}{\partial R_i}+A_0D_i(D_0U-1)+D_i(D_0U+1)\approx0.
\end{equation}

For simplicity, we denote $y_0\equiv D_0U$, $\widetilde{B}_i^\mu\equiv\sum_j y_{ij}Q_j^\mu$, $z_i\equiv D_i$, and $w_i\equiv A_0D_i$. Equations (\ref{Eq:IBP3:a}) and (\ref{Eq:IBPB2}) give rise to polynomial equations in $y$, $z$, and $w$. Since $z$ and $w$ do not commute with $y$, we assume that $z$ and $w$ are always to the left of $y$. The noncommutativity problem can be solved by avoiding multiplying $y$ by $z$ or $w$ from the right-hand side. When we try to generate a Gr\"obner basis by using the Buchberger algorithm \cite{Buchberger:1998aa}, terms of the form $y_iz_j$ or $y_iw_j$ may arise, which are in contradiction with the ordering we use. In order to avoid this kind of terms, we multiply monomials free of $z$ and $w$ by an auxiliary variable $z_0$, and add to the polynomial equation system following equations:

\begin{equation}\label{Eq:AuxPolEquat}
z_iz_j=w_iw_j=z_iw_j=0.
\end{equation}

\noindent Terms of the form $y_iz_j$ (or $y_iw_j$) arise only when we multiply an equation $f\approx0$ by $z_j$ (or $w_j$). However, the l.h.s. of the obtained equation $fz_j\approx0$ (or $fw_i\approx0$) is immediately replaced by zero due to \cref{Eq:AuxPolEquat}. (Notice that each monomial of $f$ is linear in $z$ or $w$.) Thus we can identify $y_iz_j$ (or $y_iw_j$) with $z_jy_i$ (or $w_jy_i$), since it never appears in practice. Consequently, the Gr\"obner basis for the polynomial equation system can be generated by using the Buchberger algorithm assuming all the variables are commutative. Finally we will remove polynomial equations in \cref{Eq:AuxPolEquat} from the generated Gr\"obner basis. One may use some other algorithms to generate the Gr\"obner basis. Then terms of the form $y_iz_j$ or $y_iw_j$ may not be eliminated at the intermediate steps. However, as far as we only pick those equations linear in $z$ or $w$ at the final step, the obtained basis is valid. Because all polynomial equations are homogeneous in $z$ and $w$, equations of higher degrees in $z$ and $w$ never affect those linear in $z$ or $w$. There is another type of identities. Generally the $y$ are not independent. They are related to each other through the relation

\begin{equation}\label{Eq:Temp3}
y_i=y_i(R).
\end{equation}

\noindent In practice, we first eliminate $R_i$ from \cref{Eq:Temp3}, and add the resulting equations to the polynomial equation system.

After obtaining the Gr\"obner basis, integrals of the form in \cref{Eq:TensParInt} can be reduced by reducing the polynomial $f_i$ with respect to the basis. The resulting polynomials may contain terms of the form $w_iy_jy_k\cdots I(\lambda)$. This kind of terms can be further reduced by replacing them by

\begin{equation}\label{Eq:wyRed}
w_iy_jy_k\cdots I(\lambda)=A_0\sum_a\frac{\partial y_jy_k\cdots}{\partial y_a}\left(\frac{\partial y_a(R)}{\partial R_i}I(\lambda)\right)+y_jy_k\cdots w_iI(\lambda).
\end{equation}

\noindent $A_0$ in the above equation can be replaced by its eigenvalue. $\frac{\partial y_jy_k\cdots}{\partial y_a}$ and $y_jy_k\cdots w_i$ can further be reduced with respect to the Gr\"obner basis. Terms of the form $z_iy_jy_k\cdots$ can be reduced similarly.

\section{Examples}\label{Sec:Exampl}

As an example, we consider the two-loop massless sunset diagram

\begin{equation}
I_1(-\frac{d}{2},\lambda_1,\lambda_2,\lambda_3)=\frac{e^{i\pi(\lambda_1+\lambda_2+\lambda_3)}}{(2\pi)^d}\int d^dl_1d^dl_2\frac{1}{l_1^{2(1+\lambda_1)}l_2^{2(1+\lambda_2)}(l_1+l_2+p)^{2(1+\lambda_3)}},
\end{equation}

\noindent with $p^2=1$. We first try to reduce the corresponding tensor integrals by using Method II. The polynomial $\mathcal{F}(R)$ for this integral is

\begin{equation}
\mathcal{F}_1(R)=-R_1 R_2 R_3+R_1 R_2 R_4+R_1 R_3 R_4+R_2 R_3 R_4.
\end{equation}

\noindent The operator $\widetilde{B}_i^\mu$ is

\begin{equation}
\widetilde{B}^\mu=D_0\begin{pmatrix}
R_2R_3\\
R_1R_3
\end{pmatrix}p^\mu
\equiv\begin{pmatrix}
y_2\\
y_1
\end{pmatrix}p^\mu.
\end{equation}

\noindent We define $y_3\equiv D_0U=D_0(R_1 R_2+R_1 R_3+R_2 R_3)$. Following the algorithm described in \cref{Sec:MethII}, we first generate a Gr\"obner basis out of \cref{Eq:IBPB2}. All the calculations can be done by using Mathematica. The Gr\"ober basis is generated with the built-in function \texttt{GroebnerBasis}. A degree-reverse-lexicographic order is used. To reveal the correct degree in $R_i$, we introduce another auxiliary variable $x$ and rescale the variables by $z_0\to xz_0$, and $y_i\to x^{L-1}y_i$. Experiences show that it is less efficient to generate a full basis. In practice, we exclude polynomials with degrees larger than $4$. In Mathematica, this can be implemented by representing $y_i$ by a pattern $\texttt{y[\_]}$, and setting $\texttt{y/:~y[\_]\^~n\_ = ComplexInfinity /;n>4}$. Finally, we get a basis of size $12$, among which some are (Here we have replaced the auxiliary variables $x$ and $z_0$ by $1$.)

\begin{align}
0\approx&y_3-1,\\
0\approx&y_1^2+w_2 y_3^2-w_2 y_3+z_2y_3^2+z_2y_3,\\
0\approx&2 y_2^2-w_1 y_2+w_2 y_2+w_3 y_2+w_1y_2y_3-w_2y_2y_3-w_3y_2y_3+2 w_2 y_3^2-w_1 y_1+w_2 y_1-w_3 y_1\nonumber\\
&-2 w_2 y_3+w_1 y_1 y_3-w_2 y_1 y_3+w_3 y_1 y_3+z_1y_2y_3+z_1y_2-z_2y_2y_3-z_2y_2-z_3y_2y_3-z_3y_2\nonumber\\
&+z_1y_1+z_1y_1y_3+2z_2y_3^2-z_2y_1-z_2y_1 y_3+2z_2y_3+z_3y_1+z_3y_1y_3-y_2+y_1.
\end{align}

\noindent The generated basis is complete in the sense that a monomial $y_1^{i_1}y_2^{i_2}y_3^{i_3}$ can be reduced with respect to the Gr\"obner basis as far as one of the following conditions is satisfied: $i_1>1$, $i_2>1$, or $i_3>0$. The polynomial reduction can be carried out by using the built-in function \texttt{PolynomialReduce}.

Though the basis is complete in this example, the reduction is not. The reason for this is that when we use \cref{Eq:wyRed} to reduce terms of the form $w_iy_jy_k\ldots$, we get terms of the form $\frac{\partial y_j(R)}{\partial R_i}$, which is not a polynomial in the $y$. The unreduced integrals can further be reduced by applying symbolic rules. Symbolic rules are derived by solving symbolic IBP identities generated by \cref{Eq:DimShiftFreeIBP}. Excluding some redundant rules, and considering the symmetries under permutations of indices, we get the following rules:

\begin{subequations}
\begin{align}
I_1(i_0,i_1,i_2,i_3)=&-\frac{I(i_0,i_1-1,i_2,i_3)}{i_1 \left(i_0+i_1+1\right)}\left[\left(i_1-1\right)^2+\left(2 i_2+2 i_3+7\right) \left(i_1-1\right)+6 i_0^2+i_2^2+i_3^2\right.\nonumber\\
&\left.+\left(5 \left(i_1-1\right)+5 i_2+5 i_3+17\right) i_0+7 i_2+2 i_2 i_3+7 i_3+12\right],\qquad i_1>0,\\
I_1(i_0,i_1,i_2,i_3)=&I_1(i_0,i_2,i_1,i_3),\qquad i_2>0,\\
I_1(i_0,i_1,i_2,i_3)=&I_1(i_0,i_3,i_2,i_1),\qquad i_3>0.
\end{align}
\end{subequations}

\noindent Dimensional recurrence relations can be derived by explicitly solving IBP identities. We have

\begin{equation}
I_1(i_0,0,0,0)=\frac{(i_0+2)I_1(i_0+1,0,0,0)}{6(2i_0+3)(3i_0+4)\left(3i_0+5\right)}.
\end{equation}

\noindent Obviously these rules are complete in the sense that any integral of the form $I_1(i_0,i_1,i_2,i_3)$ can be reduced to the master integral $I(-\frac{d}{2},0,0,0)$ by applying these rules.

Alternatively, we can do the reduction by using Method I. Since the first equation in \cref{Eq:abDef} has no solution, we need to extend the polynomial $\mathcal{F}_1$ by introducing two auxiliary parameters, which is equivalent to introducing two auxiliary propagators $l_1\cdot p$ and $l_2\cdot p$. We denote

\begin{equation}
\begin{split}
I_2(-\frac{d}{2},\lambda_1,\lambda_2,\lambda_3,\lambda_4,\lambda_5)=&\frac{e^{i\pi\sum_{i=1}^5\lambda_i}}{(2\pi)^d}\int d^dl_1d^dl_2\\
&\times\frac{1}{l_1^{2(1+\lambda_1)}l_2^{2(1+\lambda_2)}(l_1+l_2+p)^{2(1+\lambda_3)}(l_2\cdot p)^{1+\lambda_4}(l_1\cdot p)^{1+\lambda_5}}.
\end{split}
\end{equation}

\noindent The polynomials $A$, $B$, $\bar{A}$ and $\bar{B}$ for this integral are

\begin{align}
A=&
\begin{pmatrix}
R_1+R_3 & R_3\\
R_3 & R_2+R_3
\end{pmatrix},\\
B^\mu=&
\begin{pmatrix}
R_3+\frac{1}{2}R_5\\
R_3+\frac{1}{2}R_4
\end{pmatrix}
p^\mu,\\
\bar{A}=&\frac{1}{2}
\begin{pmatrix}
-2D_5^2-2D_1 & -2D_4D_5+D_1+D_2-D_3+2D_4+2D_5-1\\
-2 D_4 D_5+D_1+D_2-D_3+2 D_4+2 D_5-1 & -2D_4^2-2D_2
\end{pmatrix},\\
\bar{B}^\mu=&
\begin{pmatrix}
D_5\\
D_4
\end{pmatrix}
p^\mu.
\end{align}

\noindent Then tensor integrals can be parametrized by using the method described in \cref{Sec:MethI}. For example, we have

\begin{align}
&\frac{1}{(2\pi)^d}\int d^dl_1d^dl_2\frac{l_1^\mu l_2^\nu}{l_1^2l_2^2(l_1+l_2+p)^2l_2\cdot pl_1\cdot p}\\
=&\left[\bar{B}_1^\mu\bar{B}_2^\nu-\frac{1}{2}\widetilde{A}_{12}(g^{\mu\nu}-p^\mu p^\nu)\right]I_2(-\frac{d}{2},0,0,0,-1,-1)\\
=&\left[\bar{B}_1^\mu\bar{B}_2^\nu+\frac{1}{d-1}\bar{A}_{12}(g^{\mu\nu}-p^\mu p^\nu)\right]I_2(-\frac{d}{2},0,0,0,-1,-1)\\
=&p^\mu p^\nu I_2(-\frac{d}{2},0,0,0,-2,-2)+\frac{1}{2(d-1)}(g^{\mu\nu}-p^\mu p^\nu)\left[-2I_2(-\frac{d}{2},0,0,0,-2,-2)\right.\\
&+I_2(-\frac{d}{2},-1,0,0,-1,-1)+I_2(-\frac{d}{2},0,-1,0,-1,-1)-I_2(-\frac{d}{2},0,0,-1,-1,-1)\\
&\left.+2I_2(-\frac{d}{2},0,0,0,-2,-1)+2I_2(-\frac{d}{2},0,0,0,-1,-2)-I_2(-\frac{d}{2},0,0,0,-1,-1)\right]
\end{align}

\noindent It is easy to check that this result is consistent with the one obtained by using the standard tensor-reduction method. The resulting scalar integrals can further be reduced by solving IBP identities generated by using \cref{Eq:DimShiftFreeIBPII}.

As a less trivial example, we consider the reduction of the rank-4 massless double-box integral,

\begin{equation}
\frac{1}{\pi^d}\int d^dl_1d^dl_2\frac{l_1^\mu l_1^\nu l_2^\alpha l_2^\beta}{l_1^2l_2^2(l_1+k_1)^2(l_1-k_2)^2(l_1+l_2+k_1)^2(l_1+l_2-k_2)^2(l_1+l_2-k_2-k_3)^2},
\end{equation}

\noindent with $k_i^2=0,~i=1,~2,~3$, and $(k_1+k_2+k_3)^2=0$. This integral can easily be parametrized by using \cref{Eq:ParTensInt,Eq:TensGen}. The highest degree of the resulting parametric integrals is $8$. (The highest degree of the scalar integrals obtained by using the traditional tensor-reduction method is $4$.) However, the degrees can be reduced by using Method II. Following the algorithm described in \cref{Sec:MethII}, we get a Gr\"obner basis of size 46 for the top topology within a few seconds. Similar to the case of the first example, polynomials with degrees larger than $4$ are excluded from the basis. After applying this basis, the highest degree of the resulting integrals becomes $3$, which can further be (partially) reduced by applying symbolic rules.

Contrary to the first example, the generated symbolic rules for the double-box integral are incomplete. We denote a scalar integral by

\begin{equation}
\begin{split}
I_3(-\frac{d}{2},\lambda_1,\ldots,\lambda_7)\equiv&\frac{e^{i\pi\sum_{i=1}^7\lambda_i}}{\pi^d}\int
d^dl_1d^dl_2\frac{1}{l_1^{2(\lambda_1+1)}l_2^{2(\lambda_2+1)}(l_1+k_1)^{2(\lambda_3+1)}(l_1-k_2)^{2(\lambda_4+1)}}\\
&\times\frac{1}{(l_1+l_2+k_1)^{2(\lambda_5+1)}(l_1+l_2-k_2)^{2(\lambda_6+1)}(l_1+l_2-k_2-k_3)^{2(\lambda_7+1)}}.
\end{split}
\end{equation}

\noindent Integrals that cannot be reduced by applying the obtained symbolic rules are those with $\lambda_4=\lambda_5=\lambda_6=\lambda_7=0$. These unreduced integrals can easily be reduced by solving parametric IBP identities. The number of independent IBP identities to be solved is much smaller than that in the traditional momentum-space IBP method. This is because the number of unreduced scalar integrals in the former is much smaller. For example, since only those integrals with $\lambda_4=\lambda_5=\lambda_6=\lambda_7=0$ cannot be reduced by applying the symbolic rules, there are only dozens of unreduced integrals with degree $5$ in the top topology. However, there are more than one thousand momentum-space integrals with degree $5$.

All the above calculations are done by using a Mathematica code, which takes about one hour in total on a laptop (with two kernels). As a comparison, the helicity-amplitude methods \cite{Chen:2019wyb,Peraro:2019cjj} obviously break down for this example, since it is not a full amplitude. And we fail to carry out the tensor reduction by using the standard projector method on the same machine due to the lack of memory (about $2$ GiB). So, instead, we first carry out the tensor reduction by using Method I described in \cref{Sec:MethI}, and do the IBP reduction with the regular IBP method, which takes about $400$ seconds by using \texttt{FIRE6}\cite{Smirnov:2019qkx}. We fail to carry out the reduction by using the Mathematica version of \texttt{FIRE}  after a running of several hours (due to the lack of memory too).

To validate the algorithm, we reduced the amplitudes for the QCD corrections to the Higgs two-photon decay by using both methods. The result was consistent with the one obtained by combining the tensor reduction with the traditional IBP method.

\section{Summary}

In this paper, the reduction of tensor integrals is considered. Following the method developed in paper I, a tensor integral is parametrized by applying a sum of chains of operators $\widetilde{A}$ and $\widetilde{B}$ to a scalar integral (cf. \cref{Eq:ParTensInt}). Because the $\widetilde{B}$ are of degree $L$, the resulting parametric integrals are of high degrees for multiloop high-rank tensor integrals. We show that polynomial equations for the $\widetilde{B}$ can be constructed (cf. \cref{Eq:IBPB}). Based on these equations, two methods to reduce the degrees of parametric integrals are provided.

In the first method, by introducing some auxiliary parameters, we show that the $\widetilde{B}$ can be traded by the $\bar{B}$ (cf. \cref{Eq:BtSol,Eq:TensGen2}), and the $\widetilde{A}$ can be traded by the $\bar{A}$ (cf. \cref{Eq:AtChainSol}). The corresponding parametric integrals are of much lower degrees. The parametrizing result is consistent with the one obtained by using the tensor reduction method. However, the former is much easier to carry out for high-rank tensor integrals. The resulting parametric integrals can be reduced by solving dimensional-shift-free IBP identities (cf. \cref{Eq:DimShiftFreeIBPII}).

In the second method, a Gr\"obner basis is generated out of these polynomial equations of $\widetilde{B}$. Tensor integrals are partially reduced by using this basis. The unreduced integrals can further be reduced by solving parametric IBP identities combining with the application of symbolic rules. By virtue of the positivity of the indices, symbolic rules can easily be generated. The number of independent parametric IBP identities to be solved is much smaller than that in the regular IBP method.

\section*{\normalsize{Acknowledgments}}

This work was supported by the Natural Sciences and Engineering Research Council of Canada.

\appendix

\section{Derivation of eq. (\ref{Eq:IBPB})}\label{App:DerEqIBPB}

For simplicity, we use a center dot to denote both the inner product of two vectors and the product of two matrices. By using the identity $\frac{\partial A^{-1}}{\partial R_i}=-A^{-1}\cdot\frac{\partial A}{\partial R_i}\cdot A^{-1}$, \cref{Eq:IBP3} leads to

\begin{equation}
\begin{split}
D_0\frac{\partial U}{\partial R_i}R_{n+1}-D_i\approx&-D_0\frac{\partial F}{\partial R_i}\\
=&D_0\frac{\partial U}{\partial R_i}(C-B\cdot A^{-1}\cdot B)+D_0U\left(B\cdot A^{-1}\cdot \frac{\partial A}{\partial R_i}\cdot A^{-1}\cdot B-2\frac{\partial B}{\partial R_i}\cdot A^{-1}\cdot B+\frac{\partial C}{\partial R_i}\right)\\
\approx&-D_0FD_0\frac{\partial U}{\partial R_i}+\widetilde{B}\cdot\frac{\partial A}{\partial R_i}\cdot\widetilde{B}-2\frac{\partial B}{\partial R_i}\cdot\widetilde{B}+\frac{\partial C}{\partial R_i}.
\end{split}
\end{equation}

\noindent In the last step of the above equation, we have replaced $D_0U$ by $1$ or vice versa. For the first term in the last line, we have

\begin{equation}
\begin{split}
-D_0FD_0\frac{\partial U}{\partial R_i}=&-D_0(\mathcal{F}-UR_{n+1})D_0\frac{\partial U}{\partial R_i}\\
\approx&(A_0-1)D_0\frac{\partial U}{\partial R_i}+R_{n+1}D_0\frac{\partial U}{\partial R_i}\\
=&(A_0-1)D_0\frac{\partial U}{\partial R_i}+D_0\frac{\partial U}{\partial R_i}(R_{n+1}+2)\\
=&(A_0+1)D_0\frac{\partial U}{\partial R_i}+D_0\frac{\partial U}{\partial R_i}R_{n+1}\\
=&D_0A_0\frac{\partial U}{\partial R_i}+D_0\frac{\partial U}{\partial R_i}R_{n+1}.
\end{split}
\end{equation}

\noindent Combining the above equations we get

\begin{equation}
\widetilde{B}\cdot\frac{\partial A}{\partial R_i}\cdot\widetilde{B}-2\frac{\partial B}{\partial R_i}\cdot\widetilde{B}+D_0A_0\frac{\partial U}{\partial R_i}+\frac{\partial C}{\partial R_i}+D_i\approx0.
\end{equation}

\section{Proof of eq. (\ref{Eq:TensGen2})}\label{App:ProofEqTensGen2}

We first define

\begin{equation}
B_i^{\prime\mu}(q^\prime)\equiv \left(\frac{\partial}{\partial q_{i\mu}^\prime}+\bar{B}_i^\mu+\frac{1}{2}\sum_{i^\prime}\tilde{A}_{ii^\prime}q_{i^\prime}^{\prime\mu}\right),
\end{equation}

\noindent where $q^\prime$ are vectors defined in the linear space generated by the external momenta $Q$. We will show that

\begin{equation}\label{Eq:AppTemp1}
\widetilde{B}_{i_n}^{\mu_n}\cdots\widetilde{B}_{i_2}^{\mu_2}\widetilde{B}_{i_1}^{\mu_1}\approx\left[B_{i_n}^{\prime\mu_n}(q^\prime)\cdots B_{i_2}^{\prime\mu_2}(q^\prime)B_{i_1}^{\prime\mu_1}(q^\prime)\right]_{q^\prime=0}.
\end{equation}

\noindent This equation can easily be proved by iteration. By virtue of \cref{Eq:BtSol}, it holds when $n=1$. Suppose that this equation holds for $n=N$, then we have

\begin{equation}
\begin{split}
&\left[B_{i_{N+1}}^{\prime\mu_{N+1}}(q^\prime)\cdots B_{i_2}^{\prime\mu_2}(q^\prime)B_{i_1}^{\prime\mu_1}(q^\prime)\right]_{q^\prime=0}\\
=&\bar{B}_{i_N}^{\mu_{N+1}}\left[B_{i_{N}}^{\prime\mu_{N}}(q^\prime)\cdots B_{i_2}^{\prime\mu_2}(q^\prime) B_{i_1}^{\prime\mu_1}(q^\prime)\right]_{q^\prime=0}\\
&+\frac{1}{2}\sum_{j=1}^n\widetilde{A}_{j,N+1}\left[B_{i_{N}}^{\prime\mu_{N}}(q^\prime)\cdots B_{i_{j-1}}^{\prime\mu_{j-1}}(q^\prime)B_{i_{j+1}}^{\prime\mu_{j+1}}(q^\prime)\cdots B_{i_1}^{\prime\mu_1}(q^\prime)\right]_{q^\prime=0}\\
\approx&\widetilde{B}_{i_N}^{\mu_N}\cdots\widetilde{B}_{i_2}^{\mu_2}\widetilde{B}_{i_1}^{\mu_1}\bar{B}_{i_{N+1}}^{\mu_{N+1}}+\frac{1}{2}\sum_{j=1}^n\left(\widetilde{B}_{i_{N}}^{\mu_{N}}\cdots\widetilde{B}_{i_{j-1}}^{\mu_{j-1}}\widetilde{B}_{i_{j+1}}^{\mu_{j+1}}\cdots\widetilde{B}_{i_1}^{\mu_1}\widetilde{A}_{j,N+1}\right)\\
\approx&\widetilde{B}_{i_{N+1}}^{\mu_{N+1}}\cdots\widetilde{B}_{i_2}^{\mu_2}\widetilde{B}_{i_1}^{\mu_1}.
\end{split}
\end{equation}

\noindent In the last line we have used the commutation relation \cref{Eq:BtBbComm2}. Then, by using \cref{Eq:AppTemp1}, we have

\begin{equation}
P_i^\mu\approx-\frac{\partial}{\partial p_{i,\mu}}-\frac{\partial}{\partial q_{i\mu}^\prime}-\bar{B}_i^\mu+\frac{1}{2}\sum_{j=1}^L\widetilde{A}_{ij}(p_j^\mu-q_j^{\prime\mu}).
\end{equation}

\noindent We split $p$ into two parts $\bar{p}$ and $p^\prime$, such that $p^\prime$ lies in the linear space generated by $Q$ and $\bar{p}$ is orthogonal to $Q$. We introduce a new set of momenta $k=\frac{1}{2}(p^\prime+q^\prime)$, and $k^\prime=p^\prime-q^\prime$. Then it is easy to get

\begin{equation}
P_i^\mu\approx-\frac{\partial}{\partial\bar{p}_{i,\mu}}-\frac{\partial}{\partial k_{i\mu}}-\bar{B}_i^\mu+\frac{1}{2}\sum_{j=1}^L\widetilde{A}_{ij}(\bar{p}_j^{\mu}+k_j^{\prime\mu}).
\end{equation}

\noindent Because $\frac{\partial}{\partial k_{i\mu}}$ commutes with $k_j^{\prime\mu}$ and finally we will take $k_i^\mu=k_i^{\prime\mu}=0$, these two terms do not contribute. Thus they can be omitted. Hence
\begin{equation}
P_i^\mu\approx-\frac{\partial}{\partial\bar{p}_{i,\mu}}-\bar{B}_i^\mu+\frac{1}{2}\sum_{j=1}^L\widetilde{A}_{ij}\bar{p}_j^{\mu}.
\end{equation}

\section{Derivation of eq. (\ref{Eq:AtSol})}\label{App:DerEqAtSol}

We first consider the first term in \cref{Eq:IBPBa}. By using \cref{Eq:BtSol} and the commutation relation \cref{Eq:BtBbComm2}, we have

\begin{equation}
\begin{split}
\sum_{j,k}\frac{\partial A_{jk}}{\partial a_i}\widetilde{B}_{j}\cdot\widetilde{B}_{k}\approx&\sum_{j,k}\frac{\partial A_{jk}}{\partial a_i}\bar{B}_{j}\cdot\widetilde{B}_{k}\\
=&\sum_{j,k}\frac{\partial A_{jk}}{\partial a_i}\widetilde{B}_{j}\cdot\bar{B}_{k}+\frac{1}{2}\sum_{j,k,u}\delta^{uu}\frac{\partial A_{jk}}{\partial a_i}\widetilde{A}_{jk}\\
\approx&\sum_{j,k}\frac{\partial A_{jk}}{\partial a_i}\bar{B}_{j}\cdot\bar{B}_{k}+\frac{E}{2}\sum_{j,k,u}\frac{\partial A_{jk}}{\partial a_i}\widetilde{A}_{jk}
\end{split}
\end{equation}

\noindent By virtue of the Laplace expansion of $U$ (as the determinant of $A$), the second term in \cref{Eq:IBPBa} becomes

\begin{equation}
D_0A_0\frac{\partial U}{\partial a_i}=\sum_{j,k}\frac{\partial A_{jk}}{\partial a_i}\widetilde{A}_{jk}A_0.
\end{equation}

\noindent Substituting the above two equations into \cref{Eq:IBPBa}, and multiplying both sides of the obtained equation by $\alpha_{jk,i}$, we get

\begin{equation}
\bar{B}_j\cdot\bar{B}_k+\widetilde{A}_{jk}\left(A_0+\frac{E}{2}\right)+\sum_k\alpha_{jk,i}\left(D_{a_i}+\frac{\partial C}{\partial a_i}\right)\approx0.
\end{equation}

\section{Proof of eq. (\ref{Eq:AtChainSol})}\label{App:DerEqAtChainSol}

We prove \cref{Eq:AtChainSol} by iteration. Obviously it holds when $n=1$, since in this case it is just \cref{Eq:AtSol}. Now suppose that this equation holds when $n=m$. Using \cref{Eq:AtSol,Eq:AbAtComm}, we get

\begin{equation}
\begin{split}
&\Delta_{i_1i_2\cdots i_m}^{j_1j_2\cdots j_m}\widetilde{A}_{i_{m+1}j_{m+1}}\\
\approx&\widetilde{A}_{i_2j_2}\widetilde{A}_{i_3j_3}\cdots\widetilde{A}_{i_mj_m}\bar{A}_{i_1j_1}\widetilde{A}_{i_{m+1}j_{m+1}}\\
=&\widetilde{A}_{i_2j_2}\widetilde{A}_{i_3j_3}\cdots\widetilde{A}_{i_mj_m}\left(\widetilde{A}_{i_{m+1}j_{m+1}}\bar{A}_{i_1j_1}-\widetilde{A}_{i_1j_1}\widetilde{A}_{i_{m+1}j_{m+1}}\right)\\
&+\frac{1}{2}\widetilde{A}_{i_2j_2}\widetilde{A}_{i_3j_3}\cdots\widetilde{A}_{i_mj_m}\left(\widetilde{A}_{i_1i_{m+1}}\widetilde{A}_{j_1j_{m+1}}+\widetilde{A}_{i_1j_{m+1}}\widetilde{A}_{i_{m+1}j_1}\right).
\end{split}
\end{equation}

\noindent Thus

\begin{equation}
\begin{split}
&\widetilde{A}_{i_2j_2}\widetilde{A}_{i_3j_3}\cdots\widetilde{A}_{i_{m+1}j_{m+1}}\bar{A}_{i_1j_1}\\
\approx&\Delta_{i_1i_2\cdots i_m}^{j_1j_2\cdots j_m}\widetilde{A}_{i_{m+1}j_{m+1}}+\widetilde{A}_{i_1j_1}\widetilde{A}_{i_2j_2}\cdots\widetilde{A}_{i_{m+1}j_{m+1}}\\
&-\frac{1}{2}\widetilde{A}_{i_2j_2}\widetilde{A}_{i_3j_3}\cdots\widetilde{A}_{i_mj_m}\left(\widetilde{A}_{i_1i_{m+1}}\widetilde{A}_{j_1j_{m+1}}+\widetilde{A}_{i_1j_{m+1}}\widetilde{A}_{i_{m+1}j_1}\right)\\
=&\Delta_{i_1i_2\cdots i_{m+1}}^{j_1j_2\cdots j_{m+1}}.
\end{split}
\end{equation}

\noindent In the last line we have used the identity $(A_0+\frac{E}{2})\widetilde{A}_{i_{m+1}j_{m+1}}=\widetilde{A}_{i_{m+1}j_{m+1}}(A_0-1+\frac{E}{2})$.

\end{document}